\documentstyle{article}
\begin{document}
\title{Heisenberg's principle: a cosmological speculation}
\author{Benito Hern\'{a}ndez--Bermejo}
\date{}
\maketitle
\setlength{\baselineskip}{15pt}
{\em Departamento de F\'{\i}sica Fundamental, Universidad Nacional de 
Educaci\'{o}n a Distancia, Apartado 60141, 28080 Madrid, Spain.}

\mbox{}

\begin{abstract}
It is shown how the uncertainty principle can provide a mechanism for the 
generation of fluctuations of very diverse scales in the early universe.
This phenomenon could account for the large-scale structure observed today.
\end{abstract}

\mbox{}


\mbox{}

\begin{flushleft}
{\bf Introduction}
\end{flushleft}
Energy and momentum conservation can be understood in classical mechanics
as a consequence of 
the existence of an inertial frame of reference for which, by definition, a
symmetry inherent in space and time holds: homoge\-neity [1]. This means that, 
in the absence of interactions, all positions of space and all instants of 
time are physically equivalent. These properties, together with isotropy of  
space, are in fact implemented in the classical theory through the structure 
of the lagrangian, which, in the case of a free particle, cannot 
depend on the position or the time, but only on the modulus of the velocity,
${\cal L}({\bf r, v},t) \; \equiv \; {\cal L}(v^{2})$. Conservation laws
are then a corollary of this structure: they are the conditions that 
${\cal L}$ must comply if these symmetry properties are satisfied. Indeed, 
this situation is not exclusive of classical mechanics: the whole classical 
theory, including infinite systems (fields), is constructed under these 
assumptions [2]. Since these conservation laws have been formalized
starting from the concept of inertial frame, we could say conversely that 
\pagebreak
nonconservation of energy and momentum would imply that an inertial frame 
does not exist or, equivalently, that space and time are not homogeneous.

In the description of quantum phenomena, the same symmetry properties 
are maintained. This is simple to see if we take into account the 
formulation of Quantum Mechanics based on path integrals [3]: in this
formalism, the probability $P(b,a)$ of going from a point ${\bf x}_{a}$ at 
a time $t_a$ to a point ${\bf x}_{b}$ at $t_b$ is given by the absolute 
square $P(b,a) = \mid K(b,a) \mid ^{2}$ of an amplitude $K(b,a)$, which is 
constructed from the action for the corresponding classical system:
\begin{equation}
    S = \int_{t_a}^{t_b} {\cal L}({\bf x, v},t) \; \mbox{d}t 
\end{equation}
However, due to the 
measurement problem, the situation is not exactly the same as in classical 
mechanics: the uncertainty principle [4,5] indicates that certain physical 
magnitudes cannot be simultaneously measured. This is the case for 
position--momentum and time--energy:
\begin{eqnarray}
                    \Delta x_{i} \Delta p_{i} & \geq & \hbar /2  \\
                    \Delta E \Delta t & \geq & \hbar /2
\end{eqnarray}
for $i = 1,2,3$. A consequence of this is that energy and momentum
do not satisfy strict conservation laws in the quantum case. This allows the
well known existence of virtual processes [2], in which the system can 
borrow a certain amount of energy during a sufficiently small period of 
time, in such a way that conservation of energy is globally, but not exactly, 
respected. Analogously, it is simple to imagine momentum virtual processes, 
in wich a particle could borrow linear momentum in an adequately small region 
of space. Following the above reasoning, the ensuing conclusion becomes 
apparent: quantum theory describes a space-time which is homogeneous in a 
coarse-grained form. The global conservation of energy and momentum is, as we 
have seen, indicative of a global homogeneity, but the possibility of local
violation of both can be formulated in terms of a locally inhomogeneous,
fluctuating, structure. 

\mbox{}

\begin{flushleft}
{\bf A Cosmological speculation}
\end{flushleft}
One of the most compelling problems of modern cosmology arose in the last
decade, when a systematic redshift survey performed by Geller and Huchra [6] 
revealed that the large-scale structure of the universe is not homogeneous. 

Several explanations have been proposed to account for such feature.
The basic idea underlying many of them is that of the amplification of
small density fluctuations in the early universe, which spread as a result 
of the gravitational instability of the homogeneous Big Bang model. The most 
standard picture comes from the application of grand unified theories (GUT's) 
to this scenario by considering adiabatic fluctuations of a purely barionic 
matter, but there are also models based on dark matter composed 
of weakly interacting massive particles [7], hot dark matter [8], cosmic 
explosions [9], the rotation of the universe in a fourth spatial dimension 
[10], etc. Some of the previous ideas have also been improved by means of the 
introduction of nonlinear gravity effects [11]. However, none of these schemes 
has reproduced well the observations, and the problem remains open.

A complication which seems to be widespread to most models is the difficulty 
in accounting for the detected lack of homogeneity at very diverse scales, 
which range from the size of galaxies to very large structures of the order 
of hundreds of Mpc: as it can be shown [12], the observed structure can 
only arise if the original fluctuations in the early universe are of 
small amplitude but embrace very different scales of length. This problem can 
be overcome if we assume that the fluctuations which originated the observed 
distribution were produced by Heisenberg's principle. The idea is simple 
after the interpretation given in the previous section: in 
the first stages after the big bang, the universe itself was a microparticle
subject to the quantum laws. The effects of the uncertainty principle 
were consequently relevant for the dynamics of the universe as a whole.
In particular, these effects were equivalent to a lack of homogeneity in
space and time. It is clear that time was far from homogeneous in an
universe under strong expansion. The interesting idea comes from the 
application of the same reasoning to position--momentum: the lack 
of homogeneity in space, which obviously might induce a counterpart in
the matter distribution. The typical size of the irregularities we observe
today should be a remnant of the size of the quantum inhomogeneity relative
to the total size of the early universe. 

We shall investigate this question through a simple model which is equivalent 
to Einstein's one based on general relativity
for an expanding universe [13]. The model consists of a spherical mass of 
high density which explodes and expands radially. Then the velocities must be  
proportional to the distance to the centre of the mass (which is at rest). 
Since the mass, which is uniformly distributed with density $\rho = M/V$, is 
very large, the process takes place under the effect of gravitational 
deceleration. For any two points 
of positions ${\bf r}_{1} = {\bf v}_{1}t$ and ${\bf r}_{2} = {\bf v}_{2}t$, 
their relative position is ${\bf d} = ({\bf v}_{1} - {\bf v}_{2})t$. Then the 
distance between points increases with time, which is just Hubble's law. The
evolution of the process is governed by the total energy $\eta$ per unit mass 
of the system (kinetic plus gravitational). The equation of motion is simply 
obtained from the expression for the gravitational energy of a sphere, which 
can be written as:
\begin{equation}
  \frac{\dot{R}^2}{R^2 c^2} - \frac{2 \eta}{R^2 c^2} = \frac{8 \pi G \rho }{3
   c^2}
  \label{gas2}
\end{equation}
where $M$ is the total mass, $R$ is the radius of the sphere and $G = 6.67 
\cdot 10^{-11} Nm^{2}kg^{-2}$. 
As noted earlier, this apparently naive model leads to a similar equation to 
the one resulting from the general relativity, the Friedmann equation [12]:
\begin{equation}
  \frac{\dot{a}^2}{a^2 c^2} + \frac{K}{a^2} = \frac{8 \pi G \rho }{3
   c^4}
   \label{tgrb}
\end{equation}
Here $a$ is an arbitrary reference length, which changes as the universe 
evolves, $\rho $ is the mass-energy density and $K$ is the curvature of the
universe. Both equations are analogous, but the energy $\eta$ is now in
correspondence with $-K$. As can be seen, if $\eta > 0$ ($K<0$) the expansion 
continues without limit (open universe), but if $\eta < 0$ ($K>0$) gravitation 
prevails and the solution consists of an expansion followed by a compression 
(closed universe). The sign of $K$ is determined [12] by $\rho$ through the 
relation $K = (\Omega - 1)\dot{a}^2$, where $\Omega = \rho / \rho_{cr}$, and 
$\rho_{cr}$ is a critical density. Since the problem of the total mass of our
universe remains unsolved, and subsequently we do not know if it is open or 
closed, we shall assume the limit value $\eta = 0$ (or $K=0$) in the  
forthcoming calculations. An advantage of this choice is that the solution of 
the Friedmann equation is particularly simple in this case, and the relevant 
features of the next analysis become apparent. We shall also briefly comment 
on the general situation $\eta \neq 0$ at the end of the article.

For the sake of simplicity, we shall deal with the conceptually simpler gas 
model of 
equation (\ref{gas2}), which provides better insight into the physics of the 
process, while all the calculations remain entirely analogous in the frame 
of equation (\ref{tgrb}). This is due to the complete similarity between 
equations (\ref{gas2}) and (\ref{tgrb}). Then, after a straightforward 
integration we find $R = (\gamma t^{2})^{1/3}$ for the solution of 
(\ref{gas2}) in the case $\eta =0$, where $\gamma = 9GM/2$. The essential 
feature here is the dependence $R \propto t^{2/3}$. 

As we stated earlier, the size of the quantum inhomogeneity associated to the 
spatial dimension is necessary in order to estimate the size of the 
fluctuations. According to our previous interpretation of Heisenberg's 
principle, such scale of inhomogeneity is provided by the associated 
uncertainty. This is logical, since for zero uncertainties we retrieve the  
classical conservation laws, namely a complete homogeneity of space and time. 
The natural progression is to calculate the uncertainty through Heisenberg's 
relation: 
\begin{equation}
   \Delta x \simeq \hbar / \Delta p = \hbar (<p^{2}> - <p>^{2})^{-1/2} \; .
   \label{dx}
\end{equation}
Here the symbols $ < > $ denote the expected value of a quantity.
If we notice that the velocity of expansion for an 
arbitrary inner point at distance $r$ of the centre is $\dot{R}(r,t) = 
\gamma ' t^{-1/3}r$, where $\gamma ' = (4GM/3R^{3})^{1/3}$, and denote the 
sphere of matter as $S$, we then have: 
\begin{eqnarray}
   <p> & = & \int_{S} \rho \dot{R} \; \mbox{d}^{3}\mbox{r} = (9GM^{4}/16 t)^{1/3} \label{ip1} \\
   <p^{2}> & = & \int_{S} \rho M \dot{R}^{2} \; \mbox{d}^{3}\mbox{r} = 16 <p>^{2}/15 \; . \label{ip2}
\end{eqnarray}
Notice that $<p>$ is the sum over $S$ of the modulus of the momentum.
Then we obtain $\Delta p = 15^{-1/2}<p> \propto t^{-1/3}$. This time 
dependence always holds, independently of the way in which we define the 
integral for the momentum, which can only involve spatial variables. From 
equation (\ref{dx}) we are finally led to the important result $\Delta{x} 
\propto t^{1/3}$. The extent of the inhomogeneity relative to the total size 
of the expanding universe is then:
\begin{equation}
    \Delta{x} / R \propto t^{-1/3}
\end{equation}
Then the size of the fluctuations induced by Heisenberg's principle can be 
arbitrarily large as $t \rightarrow 0$, and decrease with time. In particular, 
it is clear that this process embraces very diverse scales of length.

The previous calculations have been performed under the assumption of a 
universe composed of matter, since $\rho \propto R^{-3}$ in (\ref{gas2}) (or 
equivalently $\rho \propto a^{-3}$ in (\ref{tgrb})). However, in the stages 
we are interested ($t < 10^{5}$ years), the universe was dominated by 
radiation and, in fact [14], $\rho \propto R^{-4}$. When the foregoing 
calculations are repeated for this case we obtain that $R(t) \propto t^{1/2}$, 
$\Delta p \propto t^{-1/2}$, and then $\Delta x \propto t^{1/2}$. Thus for
the relative size of the irregularities we get to
\begin{equation}
      \Delta x / R = constant \;.
\end{equation}
In an expanding universe only composed of radiation Heisenberg's principle
originates fluctuations of a fixed size, which are clearly unable to 
produce the observed large-scale features.

Although we have apparently reached deadlock, this is not the case, since 
we have worked with a model which fully discards the presence of matter.
Let us consider that such a presence produces a small change in the power 
dependence of the density, namely $\rho \propto R^{\epsilon - 4}$, with
$\epsilon > 0$. We shall write for convenience $\epsilon = 8 \delta /(1+2 
\delta)$, where $\delta $ is also small and positive. Then the same 
calculations above show that $R \propto t^{\delta + 1/2}$, $\Delta p \propto 
t^{\delta -1/2}$, and consequently $\Delta{x} \propto t^{-\delta + 1/2}$. 
Thus the size of the fluctuations is:
\begin{equation}
      \Delta x / R \propto t^{-2 \delta} \; ,
      \label{f}
\end{equation}
which again is divergent in the limit $t \rightarrow 0$. In fact, we can say 
that the situation seen in the radiation scenario is unstable, since any
perturbation in the dependence of $R(t)$ due to the presence of matter 
appears amplified in (\ref{f}) and leads to time-dependent fluctuations. 

\mbox{}

\begin{flushleft}
{\bf Conclusion}
\end{flushleft}

We have seen that Heisenberg's principle seems to be a good candidate in 
order to explain the generation of a well defined kind of fluctuations in the 
early universe, a problem very poorly understood at present. 
It is worth emphasizing that these fluctuations cover a very large (in fact 
infinite) range of scales. Moreover, fluctuations are clearly weak at any 
given scale, since they only act instantaneously on such scale. These two 
properties are, as we have seen, in full accordance with the requirements of 
the present--day cosmological models.

The obvious step after the preceding calculations is the study of the 
general case $\eta \neq 0$. It can be demonstrated that, under the same 
assumptions of the previous development, the fluctuations due to the 
uncertainty principle are present for all values of $\eta$. However, since 
a complete analysis of this problem is rather lengthy, the details will be 
submitted in a future work.

\mbox{}

\begin{flushleft}
{\bf Acknowledgements}
\end{flushleft}
I would like to thank Dr. V. Fair\'{e}n for fruitful discussions and 
suggestions, and a referee for helpful comments.

\pagebreak
\setlength{\baselineskip}{14pt}
\begin{flushleft}
{\bf References}
\end{flushleft}

\begin{description}
  
  \item[\hspace*{1ex}{\rm 1}] Landau, L. D. and Lifshitz, E. M. (1960) {\em 
      Mechanics.\/} Oxford: Pergamon Press.

  \item[\hspace*{1ex}{\rm 2}] Itzykson, C. and Zuber, J. B. (1980) {\em 
      Quantum Field Theory.\/} New York: McGraw--Hill.

  \item[\hspace*{1ex}{\rm 3}] Feynman, R. P. and Hibbs, A. R. (1965) {\em 
      Quantum Mechanics and Path Integrals.\/} New York: McGraw--Hill.
  
  \item[\hspace*{1ex}{\rm 4}] Landau, L. D. and Lifshitz, E. M. (1965) {\em 
      Quantum Mechanics, Nonrelativistic Theory.\/} Oxford: Pergamon Press.

  \item[\hspace*{1ex}{\rm 5}] Cohen--Tannoudji, C., Diu, B. and Lalo\"{e}, F. 
      (1977) {\em Quantum Mechanics.\/} Vol. 1. New York: Wiley--Interscience.

  \item[\hspace*{1ex}{\rm 6}] Geller, M. J. and Huchra, J. P. (1989) Mapping  
      the Universe. {\em Science\/}, {\bf 246}, 897--903.
  
  \item[\hspace*{1ex}{\rm 7}] Kaiser, N. (1984) On the spatial correlations 
      of Abell clusters. {\em Astrophys. J.\/}, {\bf 284}(1), L9--L12.

  \item[{\rm 8}] Centrella, J. M., Gallagher, J. S., Melott, A. 
      S. and Bushouse, H. A. (1988) A case-study of large-scale structure in 
      a hot model universe. {\em Astrophys. J.\/}, {\bf 333}(1), 24--53.
  
  \item[{\rm 9}] Ikeuchi, S. (1981) Theory of galaxy formation triggered by 
      quasar explosions. {\em Publ. Astr. Soc. Jpn.\/}, {\bf 33}(2), 211--222.

  \item[{\rm 10}] Brown, W. K. (1994) A thick, rotating universe. 
      {\em Speculat. Sci. Technol.\/}, {\bf 17}(3), 186--190.
  
  \item[{\rm 11}] Shandarin, S. F. (1994) Nonlinear dynamics of the 
      large-scale structure in the universe. {\em Physica D\/}, {\bf 77}, 
      342--353.

  \item[{\rm 12}] Bertschinger, E. (1994) Cosmic structure 
      formation. {\em Physica D\/}, {\bf 77}, 354--379.
  
  \item[{\rm 13}] Gerthsen, C., Kneser, H. O. and Vogel, H. (1977) 
      {\em Physik.\/} Berlin: Sprin\-ger--Verlag.

  \item[{\rm 14}] Davies, P. (1984) {\em The Accidental 
      Universe.\/} Cambridge: Cambridge University Press.
  
\end{description}
\end{document}